\documentclass[aps,pra,twocolumn]{revtex4}

\begin{document}

\title{Locally observable conditions for the\\
successful implementation of entangling\\
 multi-qubit quantum gates}

\author{Holger F. Hofmann}
\affiliation{%
Graduate School of Advanced Sciences of Matter, Hiroshima University Hiroshima 739-8530, Japan
}%
\author{Ryo Okamoto}
\affiliation{%
Research Institute for Electronic Science, Hokkaido University,
Sapporo 060--0812, Japan
}%
\author{Shigeki Takeuchi}
\affiliation{%
Research Institute for Electronic Science, Hokkaido University,
Sapporo 060--0812, Japan
}%

\begin{abstract}
The information
obtained from the operation of a quantum gate on only two
complementary sets of input states is sufficient to
estimate the quantum process fidelity of the gate.
In the case of entangling gates, these conditions can be used
to predict the multi qubit entanglement capability from the
fidelities of two non-entangling local operations. It is then
possible to predict highly non-classical features of the
gate such as violations of local realism from the
fidelities of two completely classical input-output relations,
without generating any actual entanglement.
\end{abstract}

\maketitle

\section{Introduction}
Quantum gates are designed to perform tasks that are
far more complex than those achieved by classical
gates. To test experimental realizations of quantum
gates, it is therefore necessary to identify the
characteristic non-classical features that define the
quantum nature of the gate. In recent reports of
experimental gates, these non-classical features were
either characterized by the capability of generating
entanglement \cite{SKa03,Bri03,Gas03,Zha05} or by full
quantum process tomography \cite{Bri04,Hua04,Lan05,Kie05}.
However, the complexity of such tests will increase
rapidly as the number of qubits increases. A more intuitive
and simple method for testing quantum gate performances is
therefore desirable.

In the following we show that the quantum coherent
properties of a gate operation can be tested efficiently
by using a pair of operations on two complementary sets
of orthogonal input states \cite{Hof05a,Oka05,Hof05b}.
The classically defined fidelities of these two operations
then provide upper and lower bounds for the quantum process
fidelity. This estimate of the process fidelity can then be used to
predict the performance of the gate for other input states.
In particular, the fidelity of generating maximally
entangled outputs from product state inputs can be estimated
using only the fidelities of local non-entangling operations.
The entanglement capability of multi-qubit gates can thus
be evaluated without actually generating any entanglement.

\section{Noisy gate operations and process fidelity}

Ideal quantum gates are described by a unitary operation
$\hat{U}_{00}$, representing the effects on both the amplitude
and the phase of an arbitrary quantum state input. Using the
computational basis $\{\mid\! n_z \rangle\}$ for the input states,
this ideal operation can be defined by
\begin{equation}
\label{eq:U}
\hat{U}_{00} \mid\! n_z \rangle = \mid\! f_n \rangle.
\end{equation}
However, any experimental implementation will include noise effects
due to decoherence. To classify these noise effects, it is
useful to expand them into a set of orthogonal unitary operations.

For $N$ qubits, this is easily accomplished by applying all possible
combinations of phase and amplitude errors to the input states. Specifically, a phase error is represented by the Pauli matrix
$Z$ and an amplitude error is represented
by the Pauli matrix $X$. The set of local operations applied to
the input are then given by the single qubit-operations
$I$, $X$, $Z$, and $ZX=iY$.
The multi-qubit error can therefore be represented by a product
of phase and amplitude errors,
$\hat{\Pi}_{ij}=\hat{\Phi}_i \hat{A}_j$,
where $\hat{\Phi}_i$ is a product operator of $N$ local $I$
and $Z$ operators describing the distribution of phase errors
and $\hat{A}_j$ is an product operator of $N$ local $I$ and $X$
operators describing the distribution of amplitude errors.
The indices $i$ and $j$ can be given by $N$-bit binary
numbers, so that the location of the errors are given by
the digits with a value of 1. With this definition of error
operators, a set of $4^{N}$ orthogonal operations
representing the possible errors of $\hat{U}_{00}$ can be
constructed by
\begin{eqnarray}
\label{eq:errors}
\hat{U}_{ij} &=& \hat{U}_{00} \hat{\Pi}_{ij},
\nonumber \\
\mbox{with} && \mbox{Tr}\{\hat{U}_{ij}^\dagger \hat{U}_{kl}\}
= 2^N \delta_{ik}\delta_{jl}.
\end{eqnarray}
Any operator can now be expressed as a linear combination of
the unitary operations $\hat{U}_{ij}$.

In general, a noisy experimental process is described by a
superoperator $E(\hat{\rho})$ that transforms
an arbitrary input density matrix $\hat{\rho}_{\mbox{in}}$ into
the corresponding output density matrix $\hat{\rho}_{\mbox{out}}
=E(\hat{\rho}_{\mbox{in}})$. This superoperator can be written
as a process matrix with matrix elements $\chi_{ij,kl}$ by
expanding it in terms of the orthogonal unitary operations
$\hat{U}_{ij}$,
\begin{equation}
\label{eq:pmat}
E(\hat{\rho}) = \sum_{i,j,k,l} \chi_{ij,kl}
\hat{U}_{ij} \hat{\rho} \hat{U}_{kl}^\dagger.
\end{equation}
The process fidelity is then defined as the overlap of the
process matrix with the ideal process $\hat{U}_{00}$,
given by the matrix element $\chi_{00,00}=F_{\mbox{process}}$,
and the diagonal elements $\chi_{ij,ij}$ of the process matrix
can be interpreted as the probabilities of the phase and
amplitude errors $ij$.

\section{Observable effects of quantum errors}

The classical operation performed by a quantum gate is tested
by preparing input states in the computational basis states
$\mid \! n_z \rangle$ and measuring the fidelities of the
predicted output states $\mid \! f_n \rangle$. The classical
fidelity of this operation is equal to the average probability
of obtaining the predicted results for all $2^N$ possible
inputs,
\begin{eqnarray}
\lefteqn{F_z = \frac{1}{2^N} \sum_n p(f_n|n_z)}
\nonumber \\ &=&
\frac{1}{2^N} \sum_n \langle f_n \! \mid
E(\mid \! n_z \rangle\langle n_z \! \mid)
\mid \! f_n \rangle.
\end{eqnarray}
Using the definition of the output state given in eq. (\ref{eq:U}),
the definition of the errors in eq. (\ref{eq:errors}), and
the definition of the process matrix in eq. (\ref{eq:pmat}),
it can be shown that the classical fidelity is given by
\begin{equation}
\label{eq:Fz}
F_z = \sum_{i} \chi_{i0,i0}.
\end{equation}
The classical fidelity $F_z$ can therefore be identified with
the sum of the $2^N$ diagonal elements
of the process matrix corresponding to phase errors only.

Eq.(\ref{eq:Fz}) is a quantitative expression of the fact that
the classical fidelity $F_z$ is not sensitive to phase errors
in the Z-basis inputs. In order to obtain information about
these phase errors, it is therefore necessary to use a different
input basis.
As the operator representation of the errors indicate, it is
convenient to use the eigenstates of the local $X$ operators
for this purpose. This set of states $\{\mid \! n_x \rangle\}$
forms an orthogonal basis set which is complementary to the
computational basis in
the sense that the overlap between any two states taken from
the different basis sets is
\begin{equation}
|\langle n_x \mid n_z^\prime \rangle|^2 =
\left(\frac{1}{2}\right)^N.
\end{equation}
The effect of the unitary operation $\hat{U}_{00}$ on the
complementary basis states is given by
\begin{equation}
\hat{U}_{00} \mid\! n_x \rangle = \mid\! g_n \rangle.
\end{equation}
Since the input states $\mid\! n_x \rangle$ are eigenstates
of the amplitude errors $\hat{A}_j$ this operation is
not sensitive to the amplitude errors.
The classical fidelity $F_x$ of the complementary operation
is therefore given by the sum of the $2^N$ diagonal elements
of the process matrix corresponding to amplitude errors only,
\begin{equation}
F_x = \frac{1}{2^N} \sum_n p(g_n|n_x) = \sum_{j} \chi_{0j,0j}.
\end{equation}
By combining the two complementary classical fidelities,
it is therefore possible to obtain a fidelity measure that
is sensitive to all error syndromes,
\begin{equation}
F_x + F_z - 1 = \chi_{00,00}
- \sum_{i,j\neq 0}\chi_{ij,ij}.
\end{equation}
Since the only positive contribution to this fidelity
measure is the process fidelity $\chi_{00,00}=F_{\mbox{process}}$,
this measure is necessarily equal to or lower than the process
fidelity. On the other hand, each of the classical fidelities
$F_z$ and $F_x$ is necessarily equal to or greater than
the process fidelity. Therefore, it is possible to obtain
lower and upper bounds of the process fidelity with \cite{Hof05a}
\begin{equation}
F_z + F_x - 1 \;\leq\; F_{\mbox{process}}
\;\leq\; \mbox{Min}\{F_z,F_x\}.
\end{equation}
This estimate of the process fidelity provides lower limits
for all other properties of the quantum gate. It is thus
possible to predict the wide range of possible quantum operations
from the operations on only $2^{N+1}$ local input states.

\section{Application to entangling operations}

One of the essential features of quantum computation is the
generation of multi qubit entanglement. The most simple
case of $N$-qubit entanglement generation is a series of
controlled-NOT operations, where the Z-state of the first
qubit decides whether the other qubits are flipped
or not,
\begin{equation}
\hat{U}_{00}=\mid \! 0_z \rangle \langle 0_z \! \mid
\otimes I^{N-1}
+\mid \! 1_z \rangle \langle 1_z \! \mid
\otimes X^{N-1}.
\end{equation}
This operation generates a maximally entangled
$N$-qubit Greenberger-Horne-Zeilinger (GHZ) state
if the first qubit input is in an X eigenstate and the
remaining $N-1$ qubit inputs are in Z eigenstates,
\begin{eqnarray}
\label{eq:entangle}
\hat{U}_{00}\mid 0_x,0_z,0_z, \ldots \rangle
&=& \frac{1}{\sqrt{2}}(\mid 0_z,0_z,0_z, \ldots \rangle
\nonumber \\ && \hspace{0.3cm}
+ \mid 1_z,1_z,1_z, \ldots \rangle)
\nonumber \\
\hat{U}_{00}\mid 1_x,0_z,0_z, \ldots \rangle
&=& \frac{1}{\sqrt{2}}
(\mid 0_z,0_z,0_z, \ldots \rangle
\nonumber \\ && \hspace{0.3cm}
- \mid 1_z,1_z,1_z, \ldots \rangle)
\nonumber \\
&\vdots&
\end{eqnarray}
However, if all input qubits are in the Z or X basis, the operation
$\hat{U}_{00}$ produces only local Z or X basis output states.
The process fidelity $F_{\mbox{process}}$ can therefore be
estimated from the local classical fidelities $F_x$ and $F_z$.
Since the fidelity $F_{\mbox{ent.}}$ of the entanglement generating
operation (\ref{eq:entangle}) is necessarily equal to or greater
than $F_{\mbox{process}}$, the minimal fidelity of $N$-qubit
entanglement is equal to $F_z+F_x-1$.

To obtain a measure of the entanglement capability, it is
useful to consider that any state with a GHZ state component
of more than 50\% ($\langle \mbox{GHZ} \mid
\hat{\rho} \mid \mbox{GHZ} \rangle>1/2$) is an $N$-qubit
entangled state. The minimal entanglement capability $C$
can thus be defined as $2 F_{\mbox{ent.}}-1$, and the
estimate obtained from the local classical fidelities
$F_z$ and $F_x$ reads
\begin{equation}
C \geq 2 F_z + 2 F_x - 3.
\end{equation}
The multi-qubit gate is therefore capable of generating
entanglement if the average of the local classical fidelities
$F_z$ and $F_x$ is greater than $3/4$.
A pair of local non-entangling operations is thus
sufficient to demonstrate the entanglement capability
of the $N$-qubit gate.

\section{Conditions for the violation of local realism}

Since the local classical fidelities $F_z$ and $F_x$ define
a minimal amount of multi-qubit entanglement capability, it
is possible to know whether the gate operation violates
the expectations associated with local realism without
ever generating actual entanglement.

In the case of a 3 qubit
operation, the GHZ paradox is based on the correlation
\cite{Mer90}
\begin{eqnarray}
K_{\mbox{GHZ}} &=& X \otimes X \otimes X
- X \otimes Y \otimes Y
\nonumber \\ && \hspace{-0.6cm}
- Y \otimes X \otimes Y
- Y \otimes Y \otimes X.
\end{eqnarray}
For any combination of classical variables $X,Y = \pm 1$, this
correlation is equal to $\pm 2$. However, the quantum mechanical
expectation values of the operator $K_{\mbox{GHZ}}$ are $\pm 4$
- twice as high as the result of any local hidden variable theory.
Quantum states can therefore violate the GHZ inequality
$|K_{\mbox{GHZ}}|\leq 2$ by as much as a factor of two.

If the entangling operation given by eq.
(\ref{eq:entangle}) works perfectly, it generates output states
with the extremal value of $|K_{\mbox{GHZ}}|= 4$, clearly
violating local realism.
If the process fidelity is smaller than one, the minimal value
of the above correlation is obtained by assuming that all errors
result in the opposite eigenvalue of $K_{\mbox{GHZ}}$,
\begin{equation}
|K_{\mbox{GHZ}}|\geq 8 F_{\mbox{process}}-4.
\end{equation}
Local realism is therefore violated
for $F_{\mbox{process}}>3/4$.
This condition is necessarily fulfilled if the average
of the classical fidelities $F_z$ and $F_x$ exceeds
$7/8$,
\begin{equation}
\frac{1}{2}(F_z + F_x) > 7/8.
\end{equation}
An average classical fidelity greater than $7/8$ in the
two local operations on X and Z basis input qubits therefore
ensures that the operation of the three qubit gate can
generate a violation of local realism from non-entangled inputs.
Similar conditions can be derived for any number of qubits.


\section{Conclusions}

The performance of a quantum gate can be estimated efficiently
by testing only two sets of complementary input states,
e.g. the computational Z-basis and the X basis generated by
local Hadamard gates on Z-basis inputs.
By obtaining estimates of the process fidelity, it is possible
to predict the performance of the gate for any other set of
input states. As a result, the performance of entangling gates
can be estimated from the non-entangling local operations
observed in the Z and X basis. The quantum features of such
multi-qubit gates can thus be predicted from the performance of
entirely classical local operations. In particular, it is possible
to show that a gate operation violates local realism without
actually generating any entangled state.

\section*{Acknowledgements}
This work was supported in part by
Core Research for Evolutional Science and Technology,
Japan Science and Technology Agency (JST-CREST).

\end{document}